\documentclass[journal]{IEEEtran}
\usepackage{mathrsfs}
\usepackage{url,times,amsmath,color,amssymb,graphicx,epsfig,psfig,cite,psfrag,subfigure,dsfont,booktabs}
\usepackage{algorithm}
\usepackage{algorithmic}
\usepackage{bm}
\usepackage{color}
\usepackage{multicol}

\begin{document}
\title{Secure Communications in NOMA System: Subcarrier Assignment and Power Allocation}

\author{Haijun Zhang,~\IEEEmembership{Senior Member,~IEEE}, Ning Yang, Keping Long,~\IEEEmembership{Senior Member,~IEEE}, and Miao Pan,~\IEEEmembership{Senior Member,~IEEE},  George K. Karagiannidis,~\IEEEmembership{Fellow,~IEEE},   and  Victor C.M. Leung,~\IEEEmembership{Fellow,~IEEE}
\thanks{Haijun Zhang, Ning Yang, and  Keping Long are with Beijing Engineering and Technology Research Center for Convergence Networks and Ubiquitous Services, University of Science and Technology Beijing, Beijing, China (e-mail: haijunzhang@ieee.org, s20150694@xs.ustb.edu.cn, longkeping@ustb.edu.cn).

Miao Pan is with Electrical \& Computer Engineering, University of Houston, TX, USA (e-mail: mpan2@uh.edu).

George K. Karagiannidis is with Aristotle University of Thessaloniki, Thessaloniki , Greece (e-mail: geokarag@auth.gr).

Victor C.M. Leung is with the Department of Electrical and Computer Engineering, The University of British Columbia, Vancouver, BC V6T 1Z4 Canada (e-mail: vleung@ece.ubc.ca).

}} \maketitle
\begin{abstract}

Secure communication is a promising technology for wireless networks because it ensures secure transmission of information. In this paper, we investigate the joint subcarrier (SC) assignment and power allocation problem for non-orthogonal multiple access (NOMA) amplify-and-forward two-way relay wireless networks, in the presence of eavesdroppers. By exploiting cooperative jamming (CJ) to enhance the security of the communication link, we aim to maximize the achievable secrecy energy efficiency by jointly designing the SC assignment, user pair scheduling and power allocation. Assuming the perfect knowledge of the channel state information (CSI) at the relay station, we propose a low-complexity subcarrier assignment scheme (SCAS-1), which is equivalent to many-to-many matching games, and then SCAS-2 is formulated as a secrecy energy efficiency maximization problem. The secure power allocation problem is modeled as a convex geometric programming problem, and then solved by interior point methods. Simulation results demonstrate that the effectiveness of the proposed SSPA algorithms under scenarios of using and not using CJ, respectively.

\end{abstract}

\section{Introduction}
{Recently, non-orthogonal multiple access (NOMA) has been considered as a promising solution to significantly improve energy efficiency for wireless communications \cite{NOMAIEEESPL2014,NOMASpectralefficiency2017,NOMASIC2017}. The main advantage of NOMA is that it can simultaneously serves multiple users on the same subcarrier (SC) to increase the system throughput \cite{NOMAIEEESPL2014}. The concept of successive interference cancellation (SIC) at the receiver sides was applied in NOMA to address the inter-user interference. NOMA can utilize different resource allocation methods to achieve a good spectral efficiency and energy efficiency performance. In \cite{NOMAQoSIEEETrans2017}, the effective capacity with power control policy was introduced to guarantee delay quality of service (QoS) for downlink NOMA system. In \cite{NOMAFairnessIEEETrans2017}, the power allocation techniques were studied to ensure fairness under instantaneous channel state information (CSI) and average CSI in NOMA system. In \cite{NOMAvisblelightICL2017}, user grouping based on user locations was applied to reduce interference and the power allocation scheme was employed to improve sum rate for visible light communication multi-cell networks.

 Meanwhile, physical layer security has drawn much attention in wireless networks\cite{SecureTransmission2017, SecureTransmission2017v2}. Due to the broadcast nature of wireless communications, wireless transmissions are exposed to unauthorized users and vulnerable to both the jamming and eavesdropping attacks. Physical layer security is regarded as an important methodology to realize secrecy transmissions against  eavesdropping attacks\cite{WSNSecIEEEAccess2017}. Specifically, secrecy capacity can be enhanced by exploiting multiple antennas additional spatial degrees of freedom in multiple-input-multiple-output (MIMO) wiretap channel \cite{PLScapacityMIMO2011, MIMOBroadcast2014}. Furthermore, researchers applied robust beamforming transmission technique, artificial noise (AN), and Multi-antenna relay scheme to improve physical layer security \cite{PLSbeamforming2016, MIMOBeamforming2012, PLSAN2017, PLSRelay-aided2015}. Additionally, the physical layer security of cooperative communication in large-scale cognitive radio networks was investigated \cite{PLSRelayCognitiveRadio2016} with invoking a multiphase transmission scheme.

Cooperative jamming (CJ) is a special physical layer technique, which uses AN to confuse the eavesdropper. In \cite{CJuntrustedrelay2012}, CJ nodes was studied and interfered untrusty relay nodes with splitted power. In \cite{CJPrecodingTwo-wayrelay2017}, the distinct precoding vectors of node users and jamming signals were designed in two-way relaying wiretap systems. In \cite{CJsignalSuperposition2017}, secure transmission schemes were designed for relay network by exploiting CJ and signal superposition methods in two typical communication scenarios. One scenario is to maintain a satisfactory transmission rate while minimizing message leakage. The other scenario is to improve the average throughput of the system. In addition, an uncoordinated cooperative jamming scheme was investigated and uncoordinated single-antenna users independently transmit jamming signal. The central control properly allocated the jamming power of each helper to optimize secrecy sum rate \cite{CJUncoordinated2017}.

Resource allocation plays a crucial role in exploiting the potential performance gain for NOMA wireless networks. Several works have employed different optimization methods to improve the sum rate in several research works, such as the monotonic optimization \cite{momotonicopt2016}, Lagrangian duality theory \cite{LagrangianDualityTheory2016}, and matching theory \cite{MatchingGame2017}. Besides the maximization of sum rate, resource allocation with security considerations for NOMA networks have also been addressed in the existing works. In \cite{SecureResource2017}, a robust resource allocation framework was investigated in half-duplex relay networks to improve the physical layer security. In \cite{SecureIEEETrans2014}, a secure cooperative communication was introduced in cognitive radio networks with users and eavesdroppers, where secondary users were allowed to access the spectrum of primary users in the presence of malicious eavesdroppers. In \cite{SecureIEEEWCL2017}, the joint relay selection and subcarrier allocation scheme was employed in decode-and-forward relay assisted secure vehicle-to-vehicle communications. However, secure resource allocation has not
been well studied for NOMA two-way relay wireless networks. These motivated our work.


A. Contribution

In this paper, we consider secrecy energy efficiency maximization based amplify-and-forward (AF) two-way relay wireless networks under power constraints. To ensure the worst-case secrecy energy efficiency for each user is a positive rate, we assume that an upper-bound capacity can be achieved for each eavesdropper. The eavesdropper is regarded as a legitimate untrusted user and the user pair patterns are known to eavesdroppers. The secure resource allocation problem is complex due to the combinatorial aspect induced by SC assignment and user pair scheduling and power allocation. Our contributions and novelties can be summarized as follows.

\hangafter 0
\hangindent 1.5em
\noindent
1) We investigate secure communications for NOMA two-way relay wireless networks in the presence of eavesdroppers under scenarios of using and not using CJ at the relay station (RS). In this work, we consider perfect CSI for the secrecy subcarrier assignment and power allocation problem and formulate it as a mixed non-convex optimization problem. The constraints include the maximum system power, the minimum data rate for each user pair, the maximum number of users can be multiplexed on the same SC and the maximum number of SCs occupied by each user pair.

\hangafter 0
\hangindent 1.5em
\noindent
2) A matching algorithm SCAS-1 is proposed for SC assignment to improve the secrecy energy efficiency. In this scheme, the user pair scheduling is updated in each iteration. The system energy efficiency improves until it converges. To achieve high secrecy energy efficiency of the system, SCAS-2 scheme is proposed in the NOMA wireless network.

\hangafter 0
\hangindent 1.5em
\noindent
3) Based on the proposed SCAS schemes, a novel power allocation scheme is proposed for the NOMA wireless network and the power allocation for each SC is derived by utilizing interior methods. To tackle this NP-hard optimization problem, the proposed SSPA-2 scheme obtains global optimal. In addition, we proposed SSPA-2 scheme to strike a balance between system performance and computational complexity.

\hangafter 0
\hangindent 1.5em
\noindent
4) Simulation results verify the derived theoretical analytical results and demonstrate the performance superiority of the proposed SSPA schemes in terms of the average secrecy without and with CJ.

B. Organization

The rest of this paper is organized as follows. Section II provides the system model and problem formulation of secure resource allocation. In Section III, secrecy energy efficiency subcarrier assignment schemes for NOMA wireless network. In Section IV, secrecy energy efficiency power allocation scheme for NOMA wireless network. In Section V, performance of the proposed algorithms are evaluated by simulations. Finally, Section VI concludes the paper.

\section{System Model}\label{sec:System Model}
\subsection{Secrecy NOMA Two-Way Relay Wireless Networks without Cooperative Jamming}\label{subsec:Secrecy Two-Way Relay without Cooperative Jamming}
We consider a NOMA two-way relay wireless network composed of $M$ preassigned user pairs, denoted by \(\mathcal{M}=\{1,\ldots,M\}\). The NOMA channel composes of $N$ SCs, denoted by \(\mathcal{N}=\{1,\ldots,N\}\), and each has a bandwidth $B$. As shown in Fig. 1, two users ($A_m$ and $B_m$), a RS, and an eavesdropper are presented. In the case of not using CJ, there is no artificial noise (AN). The bi-directional communications between users $A_m$ and $B_m$ are aided by the RS. Eavesdropper is passive and intercepts the information from $A_m$ and $B_m$ without alteration.

 AF protocol is considered in this paper which is divided into two phases: the multiple access (MA) phase and the broadcast (BC) phase. All user pairs do simultaneous wireless messages and power transfer with the RS in the MA phase; the RS further amplifies and forwards the received signals to user pairs employing its transmit power in the BC phase. Based on the CSI for NOMA two-way relay wireless network, one SC can be allocated to multiple user pairs, and one user pair can receive from the RS through multiple SCs. Meanwhile, the RS assigns different power to user pairs over SCs. Block fading channel is assumed to be flat and consisted of distance-dependent path loss and Rayleigh fading on each of the SC. We assume a slow fading environment where all the SCs are invariable during a complete transmission cycle and cochannel interference among user pairs on each SC is considered.

\begin{figure}[t]
        \centering
        \includegraphics*[width=80mm]{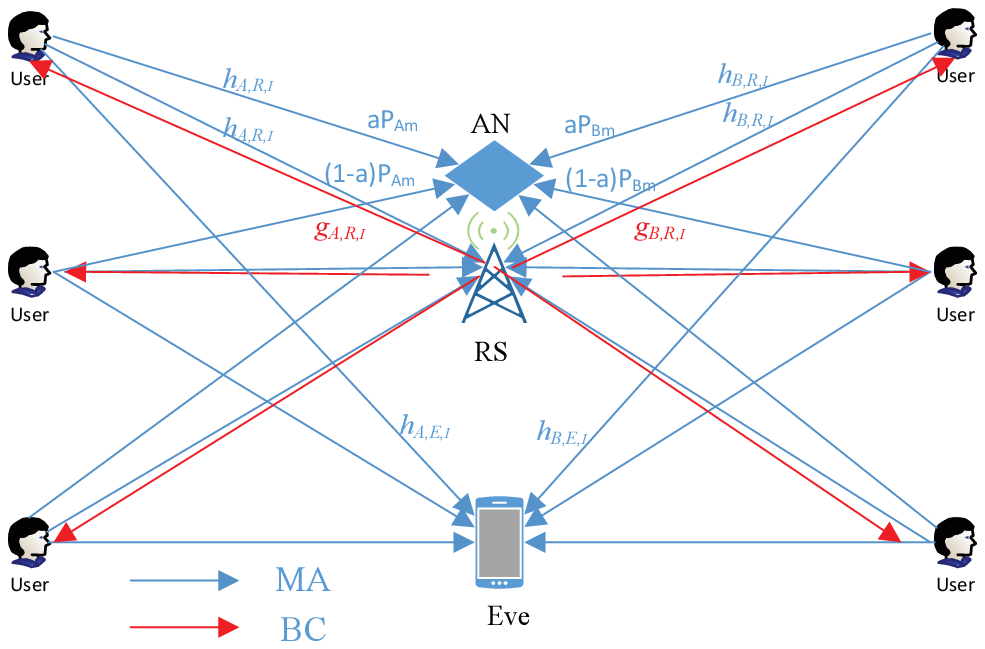}
        \caption{System model of the security transmission in two-way relay wireless network.}
        \label{fig:1}
\end{figure}

In the MA phase, we assume that SC $i$ is allocated to the $K$ user pairs, where \(\mathcal{K}=\{1,\ldots,K\}\). The $m$th user pair is composed of user $A_m$ and user $B_m$, where $m\in \mathcal K$, $m\in \mathcal M$. The received signal on SC $i$ at the RS  is given by
\begin{equation}
\begin{array}{l}
{y_{RS,i}} = \sum\limits_{m \in \mathcal K} {(\sqrt {{P_{{A_m},i}}} {h_{{A_m},R,i}}{s_{{A_m},i}}} \\
 + \sqrt {{P_{{B_m},i}}} {h_{{B_m},R,i}}{s_{{B_m},i}}) + {n_{RS,i}}
\end{array}\label{eq:Am at the first phase}
\end{equation}
where $i \in \mathcal N$; $s_{A_m,i}$ and $s_{B_m,i}$ are the transmitted signals of users $A_m$ and $B_m$ on SC $i$, respectively, which are cyclic symmetric complex Gaussian (CSCG) random variables given by \(s_{A_m,i}\sim\mathcal{CN}(0,1)\) and \(s_{B_m,i}\sim\mathcal{CN}(0,1)\) separately; \(n_{RS,i}\sim\mathcal{CN}(0,\sigma^2)\) is the additive white Gaussian noise (AWGN) at the RS on SC $i$. The total transmission power of users $A_m$ and $B_m$ are constrained by ${P_{A_m}}$ and ${P_{B_m}}$, respectively; ${h_{A_m,R,i}}$ and ${h_{B_m,R,i}}$ are, respectively, the channel gains from $A_m$ to RS and from $B_m$ to RS on SC $i$.

The received signal at the eavesdropper on SC $i$ can be expressed as
\begin{equation}
\begin{array}{l}
{y_{E,i}} = \sum\limits_{m \in \mathcal K} {(\sqrt {{P_{{A_m},i}}} {h_{{A_m},E,i}}{s_{{A_m},i}}} \\
 + \sqrt {{P_{{B_m},i}}} {h_{{B_m},E,i}}{s_{{B_m},i}}) + {n_{E,i}}
\end{array}\label{eq:received at Eve first phase w/o CJ}
\end{equation}
where the channel gains on SC $i$ from $A_m$ to the eavesdropper and from $B_m$ to the eavesdropper are denoted by ${h_{A_m,E,i}}$ and ${h_{B_m,E,i}}$ separately; \(n_{E,i}\sim\mathcal{CN}(0,\sigma^2)\) is the AWGN on SC $i$ at the eavesdropper.

In the BC phase, we assume that SC $j$ is allocated to the $m$th user pair with transmitted power \(P_{R,j}\). \(\beta_{i} y_{RS,i}\) is the signal transmitted from the RS, and \(\beta_{i}=\sqrt{P_{R,j}}/\alpha_{i}\) is the amplifying coefficient. \(\alpha_{i}\) is a normalized factor denoted by \({\alpha _i} = \sqrt {\sum\limits_{m \in \mathcal K} {({P_{{A_m},i}}|{h_{{A_m},R,i}}} {|^2} + {P_{{B_m},i}}|{h_{{B_m},R,i}}{|^2}) + {\sigma ^2}} \). The total transmission power of RSs are constrained by \(\sum\limits_{j=1}^N P_{R,j}\leq P_{\rm{R}}\). The received signal on SC $j$ at $A_m$ is that \(y_{A_m,i,j}=\sqrt {P_{R,j}}g_{A_m,j}y_{RS,i}/\alpha_{i}+n_{A_m,j}\), which can be further given by
\begin{equation}
\begin{array}{l}
{y_{{A_m},i,j}} = \sum\limits_{m \in \mathcal K} {(\sqrt {{P_{R,j}}} {g_{{A_m},j}}\sqrt {{P_{{A_m},i}}} {h_{{A_m},R,i}}{s_{{A_m},i}}/{\alpha _i}} \\
 + \sqrt {{P_{R,j}}} {g_{{A_m},j}}\sqrt {{P_{{B_m},i}}} {h_{{B_m},R,i}}{s_{{B_m},i}}/{\alpha _i})\\
 + \sqrt {{P_{R,j}}} {g_{{A_m},j}}{n_{RS,i}}/{\alpha _i} + {n_{{A_m},j}}.
\end{array}\label{eq:Am at the second phase}
\end{equation}
Similarly, the received signal on SC \(j\) at $B_m$ is \(y_{B_m,i,j}=\sqrt {P_{R,j}}g_{B_m,j}y_{RS,i}/\alpha_{i}+n_{B_m,j}\), which is further given by
\begin{equation}
\begin{array}{l}
{y_{{B_m},i,j}} = \sum\limits_{m \in \mathcal K} {(\sqrt {{P_{R,j}}} {g_{{B_m},j}}\sqrt {{P_{{A_m},i}}} {h_{{A_m},R,i}}{s_{{A_m},i}}/{\alpha _i}} \\
 + \sqrt {{P_{R,j}}} {g_{{B_m},j}}\sqrt {{P_{{B_m},i}}} {h_{{B_m},R,i}}{s_{{B_m},i}}/{\alpha _i})\\
 + \sqrt {{P_{R,j}}} {g_{{B_m},j}}{n_{RS,i}}/{\alpha _i} + {n_{{B_m},j}}
\end{array}\label{eq:Bm at the second phase}
\end{equation}
where ${g_{A_m,j}}$ and ${g_{B_m,j}}$ are, respectively, the channel gains on SC \(j\) from the RS to the $m$th user pair \(A_m\) and \(B_m\); and ${n_{A_m,j}}$ and ${n_{B_m,j}}$ are separately, AWGNs at $A_m$ and $B_m$ on SC $j$ are denoted by \({n_{A_m,j}}\sim\mathcal{CN}(0,\sigma^2)\) and \({n_{B_m,j}}\sim\mathcal{CN}(0,\sigma^2)\). ${G_{{A_m},i}} \buildrel \Delta \over = |{g_{{A_m},R,i}}{|^2}/{\sigma ^2}$ and ${G_{{B_m},i}} \buildrel \Delta \over = |{g_{{B_m},R,i}}{|^2}/{\sigma ^2}$ denote the channel response normalized by noise (CRNN) of the user $A_m$ and $B_m$ separately.

The received signal at the eavesdropper on SC $j$ in the BC phase can be expressed as
\begin{equation}
\begin{array}{l}
{y_{E,i,j}} = \sum\limits_{m \in \mathcal K} {(\sqrt {{P_{R,j}}} {g_{E,j}}\sqrt {{P_{{A_m},i}}} {h_{{A_m},R,i}}{s_{{A_m},i}}/{\alpha _i}} \\
 + \sqrt {{P_{R,j}}} {g_{E,j}}\sqrt {{P_{{B_m},i}}} {h_{{B_m},R,i}}{s_{{B_m},i}}/{\alpha _i})\\
 + \sqrt {{P_{R,j}}} {g_{E,j}}{n_{RS,i}}/{\alpha _i} + {n_{E,j}}
\end{array}\label{eq:received at Eve second phase w/o CJ}
\end{equation}
where ${g_{E,j}}$ is the channel gain from the RS to the eavesdropper on SC \(j\); and ${n_{E,j}}$ is the AWGN at the eavesdropper on SC $j$, which is denoted by \({n_{E,j}}\sim\mathcal{CN}(0,\sigma^2)\).

Assuming cochannel interference among user pairs on each SC is considered. The signal-to-interference-plus-noise ratios (SINRs) of users $A_m$ and $B_m$, sharing SC $i$ in the MA phase and SC $j$ in the BC phase, which can be respectively given by
\begin{equation}
\begin{array}{l}
SN{R_{{A_m},i,j}} = \frac{{{P_{R,j}}|{g_{{A_m},j}}{|^2}{P_{{B_m},i}}|{h_{{B_m},R,i}}{|^2}/\alpha _i^2}}{{{I_{A_m}} + ({P_{R,j}}|{g_{{A_m},j}}{|^2}/\alpha _i^2 + 1){\sigma ^2}}}
\end{array} \label{eq:SNR for Am w/o CJ}
\end{equation}
and
\begin{equation}
\begin{array}{l}
SN{R_{{B_m},i,j}} = \frac{{{P_{R,j}}|{g_{{B_m},j}}{|^2}{P_{{A_m},i}}|{h_{{A_m},R,i}}{|^2}/\alpha _i^2}}{{{I_{{B_m}}} + ({P_{R,j}}|{g_{{B_m},j}}{|^2}/\alpha _i^2 + 1){\sigma ^2}}}
\end{array}. \label{eq:SNR for Bm w/o CJ}
\end{equation}
where $I_{{A_m}}$ and $I_{{B_m}}$ are given by \eqref{eq:IAm IBm without CJ} at the top of next page.

\newcounter{TempEqCnt1}                         
      \setcounter{TempEqCnt1}{\value{equation}} 
      \setcounter{equation}{7}                           
\begin{figure*}[ht]
\begin{equation}
\begin{split}
\begin{array}{l}
{I_{A_m}} = \sum\limits_{m \in \mathcal K} {({P_{R,j}}|{g_{{A_m},j}}{|^2}{P_{{A_m},i}}|{h_{{A_m},R,i}}{|^2}/\alpha _i^2 + {P_{R,j}}|{g_{{A_m},j}}{|^2}{P_{{B_m},i}}|{h_{{B_m},R,i}}{|^2}/\alpha _i^2} )\\
 - ({P_{R,j}}|{g_{{A_m},j}}{|^2}{P_{{A_m},i}}|{h_{{A_m},R,i}}{|^2}/\alpha _i^2 + {P_{R,j}}|{g_{{A_m},j}}{|^2}{P_{{B_m},i}}|{h_{{B_m},R,i}}{|^2}/\alpha _i^2) \\
{I_{{B_m}}} = \sum\limits_{m \in \mathcal K} {({P_{R,j}}|{g_{{B_m},j}}{|^2}{P_{{A_m},i}}|{h_{{A_m},R,i}}{|^2}/\alpha _i^2 + {P_{R,j}}|{g_{{B_m},j}}{|^2}{P_{{B_m},i}}|{h_{{B_m},R,i}}{|^2}/\alpha _i^2)} \\
 - ({P_{R,j}}|{g_{{B_m},j}}{|^2}{P_{{A_m},i}}|{h_{{A_m},R,i}}{|^2}/\alpha _i^2 + {P_{R,j}}|{g_{{B_m},j}}{|^2}{P_{{B_m},i}}|{h_{{B_m},R,i}}{|^2}/\alpha _i^2)
\end{array}. \label{eq:IAm IBm without CJ}
\end{split}
\end{equation}\hrulefill
\end{figure*}

\begin{figure*}[ht]
\begin{equation}
\begin{split}
\begin{array}{l}
ET = \sum\limits_{m \in \mathcal K} {({P_{{A_m},i}}|{h_{{A_m},E,i}}{|^2} + {P_{{B_m},i}}|{h_{{B_m},E,i}}{|^2})}  - ({P_{{A_m},i}}|{h_{{A_m},E,i}}{|^2} + {P_{{B_m},i}}|{h_{{B_m},E,i}}{|^2}) + {\sigma ^2}\\
ER = \sum\limits_{m \in \mathcal K} {({P_{R,j}}|{g_{E,j}}{|^2}{P_{{A_m},i}}|{h_{{A_m},R,i}}{|^2}/\alpha _i^2}  + {P_{R,j}}|{g_{E,j}}{|^2}{P_{{B_m},i}}|{h_{{B_m},R,i}}{|^2}/\alpha _i^2)\\
 - ({P_{R,j}}|{g_{E,j}}{|^2}{P_{{A_m},i}}|{h_{{A_m},R,i}}{|^2}/\alpha _i^2 + {P_{R,j}}|{g_{E,j}}{|^2}{P_{{B_m},i}}|{h_{{B_m},R,i}}{|^2}/\alpha _i^2) + ({P_{R,j}}|{g_{E,j}}{|^2}/\alpha _{m,i}^2){\sigma ^2}
\end{array} . \label{eq:ET ER without CJ}
\end{split}
\end{equation}\hrulefill
\end{figure*}

Based on \eqref{eq:received at Eve first phase w/o CJ} and \eqref{eq:received at Eve second phase w/o CJ}, the received signal in the two phases of the eavesdropper can be modeled as a \(2\)-by-\(2\) point-to-point MIMO channel expressed as
\begin{equation}
{{\bm y_E} = {\bm H_E}\bm s + {\bm n_E}} \label{eq:equivalent MIMO channel for Eve}
\end{equation}
where
\begin{equation}
\begin{array}{l}
{H_E} = \left. {\left[ {\begin{array}{*{20}{c}}
{{h_{1,{A_m}}}}&{{h_{1,{B_m}}}}\\
{{h_{2,{A_m}}}}&{{h_{2,{B_m}}}}
\end{array}} \right.} \right]\\
{h_{1,{A_m}}} = \sqrt {{P_{{A_m}}}_{,i}} {h_{{A_m},E,i}}\\
{h_{1,{B_m}}} = \sqrt {{P_{{B_m}}}_{,i}} {h_{{B_m},E,i}}\\
{h_{2,{A_m}}} = \sqrt {{P_{R,j}}} {g_{E,j}}\sqrt {{P_{{A_m},i}}} {h_{{A_m},R,i}}/{\alpha _i}\\
{h_{2,{B_m}}} = \sqrt {{P_{R,j}}} {g_{E,j}}\sqrt {{P_{{B_m},i}}} {h_{{B_m},R,i}}/{\alpha _i}
\end{array}\label{eq:equivalent channel matrix for Eve w/o CJ}
\end{equation}

\begin{equation}
s = {\left[ {\begin{array}{*{20}{c}}
{{s_{{A_m},i}}}&{{s_{{B_m},i}}}
\end{array}} \right]^T}
\end{equation}

\begin{equation}
\begin{array}{l}
{n_E} = {\left[ {\begin{array}{*{20}{c}}
{{n_{{A_m},i}}}&{{n_{{B_m},i}}}
\end{array}} \right]^T}\\
{n_{{A_m},i}} = \sum\limits_{m \in \mathcal K} {({h_{1,{A_m}}}}  + {h_{1,{B_m}}}) + {n_{E,i}} - ({h_{1,{A_m}}} + {h_{1,{B_m}}})\\
{n_{{B_m},i}} = \sum\limits_{m \in \mathcal K} {({h_{2,{A_m}}}}  + {h_{2,{B_m}}}) + \sqrt {{P_{R,j}}} {g_{E,j}}{n_{RS,i}}/{\alpha _i}\\
 + {n_{E,j}} - ({h_{2,{A_m}}} + {h_{2,{B_m}}})
\end{array}.
\end{equation}

For users \(A_m\) and \(B_m\), the instantaneous mutual information (IMI) rate are expressed as
\begin{equation}
{R_{A_m,i,j}} = \frac{1}{2}B\log (1 + SN{R_{A_m,i,j}})
\end{equation}
and
\begin{equation}
{R_{B_m,i,j}} = \frac{1}{2}B\log (1 + SN{R_{B_m,i,j}})
\end{equation}
respectively.

For the eavesdropper, due to \eqref{eq:equivalent MIMO channel for Eve} is equivalent to a \(2\)-by-\(2\) point-to-point MIMO system with the transmit signals \(s = {\left[ {\begin{array}{*{20}{c}}
{{s_{{A_1},i}}}&{{s_{{B_1},i}}}&{...}&{{s_{{A_m},i}}}&{{s_{{B_m},i}}}
\end{array}} \right]^T}\), denoted by \(\bm s\sim\mathcal{CN}(0,\bm I)\). The maximum achievable received signal for the eavesdropper is defined as \cite[Chap. 8]{tse2005fundamentals}
\begin{equation}
\begin{array}{l}
{R_{E,i,j}} = \frac{1}{2}B\log \det (\bm I + {\bm H_E}\bm H_E^H\bm Q_E^{{\rm{ - }}1})
\end{array}\label{eq:recevied signal for Eve w/o CJ}
\end{equation}
where

\begin{equation}
{Q_E} = E\left[ {{n_E}n_E^H} \right] = \left[ {\begin{array}{*{20}{c}}
{ET}&0\\
0&{ER}
\end{array}} \right]
\end{equation}\label{eq:noise convariance matrix for Eve w/o CJ}
$ET$ and $ER$ are given by \eqref{eq:ET ER without CJ} at the top of this page.
The factor $\frac{1}{2}$ and \(\mathbb E[\cdot]\) reflects the statistical average in \eqref{eq:recevied signal for Eve w/o CJ} in a complete transmission slot accounts for the two phases. Hence, the worst-case secrecy sum rate for the $m$th users over the SC pair \((i,j)\) is expressed as \cite{tekin2008general}
\begin{equation}
R_{{\rm {sec}},m,i,j}=[R_{A_m,i,j}+R_{B_m,i,j}-{R_{E,i,j}}]^+
\end{equation}
where ${[x]^ + } = \max \{ 0,x\}$.

\subsection{Secrecy NOMA Two-Way Relay Wireless Networks with Cooperative Jamming}\label{subsec:Secrecy Two-Way Relay with Cooperative Jamming}
Base on the CSI for the RS, we investigate a similar problem which is described in Section \ref{subsec:Secrecy Two-Way Relay without Cooperative Jamming} with CJ.

In the MA phase, \(s^\prime_{A_m,i}\) and \(s^\prime_{B_m,i}\) are, respectively, the incorporating jamming signals of the users $A_m$ and $B_m$ on SC $i$ for the exchange messages. For generating the portion of the transmission power of the AN are denoted by \(\alpha_{1,i}\) and \(\alpha_{2,i}\) at \(A_m\) and \(B_m\) on SC \(i\) separately. Accordingly, \(A_m\) splits its transmission power on SC \(i\) into \((1-\alpha_{1,i})P_{A_m,i}\) for exchange message \(s_{A_m,i}\), and \(\alpha_{1,i}P_{A_m,i}\) for AN, \(s^\prime_{A_m,i}\), respectively. Similar transmission scheme is used for \(B_m\). The received signal on SC $i$ at the RS  can be expressed as
\begin{equation}
\begin{array}{l}
{y_{RS,i}} = \sum\limits_{m \in \mathcal K} {(\sqrt {(1 - {\alpha _{1,i}}){P_{{A_m},i}}} {h_{{A_m},R,i}}{s_{{A_m},i}}} \\
 + \sqrt {(1 - {\alpha _{2,i}}){P_{{B_m},i}}} {h_{{B_m},R,i}}{s_{{B_m},i}}\\
 + \sqrt {{\alpha _{1,i}}{P_{{A_m},i}}} {h_{{A_m},R,i}}s{'_{{A_m},i}}\\
 + \sqrt {{\alpha _{2,i}}{P_{{B_m},i}}} {h_{{B_m},R,i}}s{'_{{B_m},i}}) + {n_{RS,i}}.
\end{array} \label{eq:receive at the RS in MA}
\end{equation}

For the eavesdropper, the received signal on SC $i$ is given by
\begin{equation}
\begin{array}{l}
y_{_E}^{(1)} = \sum\limits_{m \in \mathcal K} {(\sqrt {(1 - {\alpha _{1,i}}){P_{{A_m},i}}} {h_{{A_m},E,i}}{s_{{A_m},i}}} \\
 + \sqrt {(1 - {\alpha _{2,i}}){P_{{B_m},i}}} {h_{{B_m},E,i}}{s_{{B_m},i}}\\
 + \sqrt {{\alpha _{1,i}}{P_{{A_m},i}}} {h_{{A_m},E,i}}s{'_{{A_m},i}}\\
 + \sqrt {{\alpha _{2,i}}{P_{{B_m},i}}} {h_{{B_m},E,i}}s{'_{{B_m},i}}) + {n_{E,i}}.
\end{array} \label{eq:receive at the Eve in MA}
\end{equation}

Assuming AN signals \(s^\prime_{A_m,i}\) and \(s^\prime_{B_m,i}\), which are fully known by the RS before being transmitted over some higher-layer cryptographic protocols, so \(\sqrt {{\alpha _{1,i}}{P_{{A_m,i}}}} {h_{{A_m},R,i}}s_{{A_m},i}^{'}\) and \(\sqrt{{\alpha _{2,i}}{P_{{B_m,i}}}} {h_{{B_m},R,i}}s_{{B_m},i}^{'}\) in \eqref{eq:receive at the RS in MA} can be canceled \cite{NgSecureTWC2014}\cite{HongXingTVT2015} at the RS. As a result, only \(s_{{A_m},i}\) and \(s_{{B_m},i}\) are broadcasted on SC \(j\) in BC phase. Then, \(A_m\) and \(B_m\) can subtract \(s_{A_m}\) and subtract \(s_{B_m}\) to obtain desirable signal from the broadcast signal, respectively. However, eavesdropper suffers from large interference due to ANs are kept strictly confidential to the eavesdropper. After canceling \(s_{{A_m},i}^{'}\) and \(s_{{B_m},i}^{'}\), \(y_{RS,i}^\prime\), which transmits the remaining signal be expressed as
\begin{equation}
\begin{array}{l}
y_{_{RS,i}}^{'} = \sum\limits_{m \in \mathcal K} {(\sqrt {(1 - {\alpha _{1,i}}){P_{{A_m},i}}} {h_{{A_m},R,i}}{s_{{A_m},i}}} \\
 + \sqrt {(1 - {\alpha _{2,i}}){P_{{B_m},i}}} {h_{{B_m},R,i}}{s_{{B_m},i}}) + {n_{RS,i}}
\end{array}
\end{equation}
the amplifying coefficient is denoted by \(\beta_{i}=\sqrt{P_{R,j}}/\gamma_{i}\)
where \(\gamma_{i}\) can be regarded as a normalized factor for the forwarded signal and it is shown in \eqref{eq:gamma with CJ} at the top of the next page. The transmit power of the RS on SC \(j\) is denoted by \(P_{R,j}\). Note that we can further simplify the received signal at the \(A_m\)  by  substituting \(\beta_{i}\) and \(y_{RS,i}^\prime\) due to \(A_m\) can successfully cancel its previously transmitted \(s_{A_m,i}\) at its receiver. In the BC phase, the received signal at the \(A_m\) is given by

      \setcounter{TempEqCnt1}{\value{equation}} 
      \setcounter{equation}{21}                           
\begin{figure*}[ht]
\begin{equation}
\begin{split}
{\gamma _i} = \sqrt {\sum\limits_{m \in \mathcal K} {((1 - {\alpha _{1,i}}){P_{{A_m},i}}|{h_{{A_m},R,i}}{|^2} + (1 - {\alpha _{2,i}}){P_{{B_m},i}}|{h_{B_m,R,i}}{|^2})}  + {\sigma ^2}} .
\end{split} \label{eq:gamma with CJ}
\end{equation}\hrulefill
\end{figure*}

\begin{figure*}[ht]
\begin{equation}
\begin{split}
\begin{array}{l}
ET^{'} = \sum\limits_{m\in \mathcal M} {({\alpha _{1,i}}{P_{{A_m},i}}|{h_{{A_m},R,i}}{|^2} + {\alpha _{2,i}}{P_{{B_m},i}}|{h_{{B_m},R,i}}{|^2})}  - ({\alpha _{1,i}}{P_{{A_m},i}}|{h_{{A_m},R,i}}{|^2} + {\alpha _{2,i}}{P_{{B_m},i}}|{h_{{B_m},R,i}}{|^2}) + {\sigma ^2}\\
 + \sum\limits_{m\in \mathcal M} {((1 - {\alpha _{1,i}}){P_{{A_m},i}}|{h_{{A_m},R,i}}{|^2}}  + (1 - {\alpha _{2,i}}){P_{{B_m},i}}|{h_{{B_m},R,i}}{|^2})\\
ER^{'} = \sum\limits_{m\in \mathcal M} {({P_{R,j}}|{g_{R,E,j}}{|^2}{P_{{A_m},i}}|{h_{{A_m},R,i}}{|^2}/\gamma _i^2}  + {P_{R,j}}|{g_{R,E,j}}{|^2}{P_{{B_m},i}}|{h_{{B_m},R,i}}{|^2}/\gamma _i^2)\\
 - ({P_{R,j}}|{g_{R,E,j}}{|^2}{P_{{A_m},i}}|{h_{{A_m},R,i}}{|^2}/\gamma _i^2 + {P_{R,j}}|{g_{R,E,j}}{|^2}{P_{{A_m},i}}|{h_{{A_m},R,i}}{|^2}/\gamma _i^2) + ({P_{R,j}}|{g_{R,E,j}}{|^2}/\gamma _i^2){\sigma ^2}
\end{array} . \label{eq:ET ER with CJ}
\end{split}
\end{equation}\hrulefill
\end{figure*}

\begin{figure*}[ht]
\begin{equation}
\begin{split}
\begin{array}{l}
{I_{{A_m}}^{'}} = \sum\limits_{m \in \mathcal K} {((1 - {\alpha _{1,i}}){P_{R,j}}{P_{{A_m},i}}|{g_{R,{A_m},j}}{|^2}|{h_{{A_m},R,i}}{|^2}/\gamma _i^2 + (1 - {\alpha _{2,i}}){P_{R,j}}{P_{{B_m},i}}|{g_{R,{A_m},j}}{|^2}|{h_{{B_m},R,i}}{|^2}/\gamma _i^2} )\\
 - ((1 - {\alpha _{1,i}}){P_{R,j}}{P_{{A_m},i}}|{g_{R,{A_m},j}}{|^2}|{h_{{A_m},R,i}}{|^2}/\gamma _i^2 + (1 - {\alpha _{2,i}}){P_{R,j}}{P_{{B_m},i}}|{g_{R,{A_m},j}}{|^2}|{h_{{B_m},R,i}}{|^2}/\gamma _i^2) \\
{I_{{B_m}}^{'}} = \sum\limits_{m \in \mathcal K} {((1 - {\alpha _{1,i}}){P_{R,j}}{P_{{A_m},i}}|{g_{R,{B_m},j}}{|^2}|{h_{{A_m},R,i}}{|^2}/\gamma _i^2 + (1 - {\alpha _{2,i}}){P_{R,j}}{P_{{B_m},i}}|{g_{R,{B_m},j}}{|^2}|{h_{{B_m},R,i}}{|^2}/\gamma _i^2} )\\
 - ((1 - {\alpha _{1,i}}){P_{R,j}}{P_{{A_m},i}}|{g_{R,{B_m},j}}{|^2}|{h_{{A_m},R,i}}{|^2}/\gamma _i^2 + (1 - {\alpha _{2,i}}){P_{R,j}}{P_{{B_m},i}}|{g_{R,{B_m},j}}{|^2}|{h_{{B_m},R,i}}{|^2}/\gamma _i^2)
\end{array}. \label{eq:IAm IBm with CJ}
\end{split}
\end{equation}\hrulefill
\end{figure*}

\begin{align}
\begin{array}{l}
{y_{{A_m},i,j}} = \\
\sum\limits_{m \in \mathcal K} {(\sqrt {(1 - {\alpha _{1,i}}){P_{R,j}}{P_{{A_m},i}}} {h_{{A_m},R,i}}{g_{{A_m},R,i}}{s_{{A_m},i}}/{\gamma _i}} \\
 + \sqrt {(1 - {\alpha _{2,i}}){P_{R,j}}{P_{{B_m},i}}} {h_{{B_m},R,i}}{g_{{A_m},R,j}}{s_{{B_m},i}}/{\gamma _i})\\
 + \sqrt {{P_{R,j}}} {g_{{A_m},R,j}}{n_{RS,i}}/{\gamma _i} + {n_{{A_m},j}}.
\end{array} \label{eq:receive at Am in BC}
\end{align}
Similarly, the received signal at the \(B_m\) is defined as
\begin{align}
\begin{array}{l}
{y_{{B_m},i,j}} = \\
\sum\limits_{m \in \mathcal K} {(\sqrt {(1 - {\alpha _{1,i}}){P_{R,j}}{P_{{A_m},i}}} {h_{{A_m},R,i}}{g_{{B_m},R,i}}{s_{{A_m},i}}/{\gamma _i}} \\
 + \sqrt {(1 - {\alpha _{2,i}}){P_{R,j}}{P_{{B_m},i}}} {h_{{B_m},R,i}}{g_{{B_m},R,j}}{s_{{B_m},i}}/{\gamma _i})\\
 + \sqrt {{P_{R,j}}} {g_{{B_m},R,j}}{n_{RS,i}}/{\gamma _i} + {n_{{B_m},j}}.
\end{array}\label{eq:receive at Bm in BC}
\end{align}
For the eavesdropper, it receives a combined signal of \(s_{A_m,i}\) and \(s_{B_m,i}\) due to the transmission signals are unknown, which is expressed as
\begin{align}
\begin{array}{l}
y_{_E}^{(2)} = \\
\sum\limits_{m \in \mathcal K} {(\sqrt {(1 - {\alpha _{1,i}}){P_{R,j}}{P_{{A_m},i}}} {h_{{A_m},R,i}}{g_{R,E,j}}{s_{{A_m},i}}/{\gamma _i}} \\
 + \sqrt {(1 - {\alpha _{2,i}}){P_{R,j}}{P_{{B_m},i}}} {h_{{B_m},R,i}}{g_{R,E,j}}{s_{{B_m},i}}/{\gamma _i})\\
 + \sqrt {{P_{R,j}}} {g_{R,E,j}}{n_{RS,i}}/{\gamma _i} + {n_{E,j}}.
\end{array}\label{eq:receive at the Eve in BC}
\end{align}

Based on \eqref{eq:receive at the Eve in MA} and \eqref{eq:receive at the Eve in BC}, the received signals at the eavesdropper during a transmit slot to the equivalent point-to-point \(2\)-by-\(2\) MIMO channel can be combined into
\begin{align}
{y_E} = {H_E}^{'}s + {n_E}^{'}\label{eq:equivalent MIMO channel for Eve with CJ}
\end{align}

The equivalent channel matrix between the user pairs and the eavesdropper over the SC pair \((i, j)\) is defined as
 \begin{align}
\begin{array}{l}
{H_E} = \left. {\left[ {\begin{array}{*{20}{c}}
{{h_{1,{A_m}}}^{'}}&{{h_{1,{B_m}}}^{'}}\\
{{h_{2,{A_m}}}^{'}}&{{h_{2,{B_m}}}^{'}}
\end{array}} \right.} \right]\\
{h_{1,{A_m}}^{'}} = \sqrt {(1 - {\alpha _{1,i}}){P_{{A_m},i}}} {h_{{A_m},E,i}}\\
{h_{1,{B_m}}^{'}} = \sqrt {(1 - {\alpha _{2,i}}){P_{{B_m},i}}} {h_{{B_m},E,i}}\\
{h_{2,{A_m}}^{'}} = \sqrt {(1 - {\alpha _{1,i}}){P_{R,j}}{P_{{A_m},i}}} {h_{{A_m},R,i}}{g_{R,E,j}}\\
{h_{2,{B_m}}^{'}} = \sqrt {(1 - {\alpha _{2,i}}){P_{R,j}}{P_{{B_m},i}}} {h_{{B_m},R,i}}{g_{R,E,j}}
\end{array}\label{eq:equivalent channel matrix for Eve with CJ}
\end{align}

In the MA phase, $\tilde n_{{A_1},i}$ indicates that the equivalent received noise at the eavesdropper treating with the generated AN by the user pair as noise, which is denoted by
 \begin{align}
\begin{array}{l}
{n_{{A_m},i}^{'}} = \sum\limits_{m \in \mathcal K} {({h_{1,{A_m}}^{'}}}  + {h_{1,{B_m}}^{'}}) + {n_{E,i}} - ({h_{1,{A_m}}^{'}} + {h_{1,{B_m}}^{'}}).
\end{array}\label{eq:interference for Eve in MA}
\end{align}
In the BC phase, $\tilde{n}_2$ denotes the amplified noise and the additive noise, which is given by
 \begin{align}
\begin{array}{l}
{n_{{B_1},i}^{'}} = \sum\limits_{m \in \mathcal K} {({h_{2,{A_m}}^{'}}}  + {h_{2,{B_m}}^{'}}) - ({h_{2,{A_m}}^{'}} + {h_{2,{B_m}}^{'}})\\
 + \sqrt {{P_{R,j}}} {g_{R,E,j}}{n_{RS,i}}/{\gamma _i} + {n_{E,j}}
\end{array}. \label{eq:interference for Eve in BC}
 \end{align}
 For this equivalent noise the associated covariance matrix at the eavesdropper can be expressed as
 \begin{align}
\begin{array}{l}
{Q_E} = E\left[ {{{({n_{{A_m}}}_{,i},{n_{{B_m}}}_{,i})}^H}({n_{{A_m}}}_{,i},{n_{{B_m}}}_{,i})} \right]\\
 = \left[ {\begin{array}{*{20}{c}}
{ET}^{'}&0\\
0&{ER}^{'}
\end{array}} \right]
\end{array}. \label{eq:noise convariance matrix for Eve with CJ}
 \end{align}
where $ET^{'}$ and $ER^{'}$ are given by \eqref{eq:ET ER with CJ} at the top of this page.
 From \eqref{eq:receive at Am in BC} and \eqref{eq:receive at Bm in BC}, the SINRs of users \(A_m\) and \(B_m\), which share SC $i$ in the MA phase and SC $j$ in the BC phase, can be respectively denoted as
 \begin{align}
SN{R_{{A_m},i,j}}^{'} = \frac{{(1 - {\alpha _{2,i}}){P_{R,j}}{P_{{B_m},i}}|{g_{R,{A_m},j}}{|^2}|{h_{{B_m},R,i}}{|^2}/\gamma _i^2}}{{{I_{A_m}^{'}} + ({P_{R,j}}|{g_{R,{A_m},j}}{|^2}/\gamma _i^2 + 1){\sigma ^2}}} \label{eq:SNR for Am with CJ}
 \end{align}
 and
 \begin{align}
SN{R_{{B_m},i,j}}^{'} = \frac{{(1 - {\alpha _{1,i}}){P_{R,j}}{P_{{A_m},i}}|{g_{R,{B_m},j}}{|^2}|{h_{{A_m},R,i}}{|^2}/\gamma _i^2}}{{{I_{{B_m}}^{'}} + ({P_{R,j}}|{g_{R,{B_m},j}}{|^2}/\gamma _i^2 + 1){\sigma ^2}}}. \label{eq:SNR for Bm with CJ}
 \end{align}
where $I_{{A_m}}^{'}$ and $I_{{B_m}}^{'}$ are given by \eqref{eq:IAm IBm with CJ} at the top of this page.

The IMI rate for the user \(A_m\) and \(B_m\)  are expressed as
 \begin{align} \tilde R_{A_m,i,j}=\frac{1}{2}B\log_2(1+SNR^\prime_{A_m,i,j}) \end{align} and
 \begin{align}  \tilde R_{B_m,i,j}=\frac{1}{2}B\log_2(1+SNR^\prime_{B_m,i,j}) \end{align} respectively.

For the eavesdropper, due to \eqref{eq:equivalent MIMO channel for Eve with CJ} is equivalent to a \(2\)-by-\(2\) point-to-point MIMO system with transmission signals \(s = {\left[ {\begin{array}{*{20}{c}}
{{s_{{A_1},i}}}&{{s_{{B_1},i}}}&{...}&{{s_{{A_m},i}}}&{{s_{{B_m},i}}} \end{array}} \right]^T}\), denoted by \(\bm s\sim\mathcal{CN}(0,\bm I)\). The maximum achievable received signal for the eavesdropper is defined as \cite[Chap. 8]{tse2005fundamentals}
 \begin{align}
 \tilde R_{E,i,j}=\frac{1}{2}B\log_2\det\left(\bm I+\bm { H_{E}}\bm { H^H_{E}}\bm {\tilde Q^{-1}_{E}}\right). \label{eq:achievable rate at Eve}
 \end{align}
Hence, the worst-case secrecy sum rate with CJ for the $m$th user pair over the SC pair \((i,j)\) can be denoted by \cite{tekin2008general}
 \begin{align}
   \tilde R_{{\rm {sec}},m,i,j}=[\tilde R_{A_m,i,j}+\tilde R_{B_m,i,j}-{\tilde R_{E,i,j}}]^+. \label{eq:secrecy sum rate with CJ}
 \end{align}

\subsection{Problem Formulation}\label{subsec:Problem Formulation}
We introduce a $N \times M$ SC matrix in which the binary element ${c_{m,i,j}}$ denotes whether $m$th user pair is allocated to SC $i$ in the MA phase and SC $j$ in the BC phase. For energy efficient secure communication, our objective is to maximize the total secrecy sum rate of the system by setting the variables $\{ {c_{m,i,j}},{p_{m,j}}\}$. The energy efficiency of the system is formulated as
\begin{equation}
\begin{aligned}
{\eta _{E}}({c_{m,i,j}},{p_{m,j}}) = \frac{{{R_{\sec ,m,i,j}}({c_{m,i,j}},{p_{m,j}})}}{{{P_s}({c_{m,i,j}},{p_{m,j}})}}
\end{aligned}
\end{equation}
where ${P_s}({c_{m,i,j}},{p_{m,j}}) = {P_c} + {P_T}$, $P_T$ and $P_c$ are transmission power and the circuit power consumption, respectively.
Accordingly, the energy efficiency maximization problem is defined as

\begin{equation}\label{eq:weighted Sum rate maximization problem formulation}
\mathop {\max }\limits_{{c_{m,i,j}},{p_{m,j}}} \sum\limits_{m\in \mathcal M} {\sum\limits_{i \in \mathcal N} {\sum\limits_{j \in \mathcal N} {{\eta _{E}}({c_{m,i,j}},{p_{m,j}})} } }
\end{equation}

\begin{equation}
\begin{aligned}
& \text{subject to}
& & {C1:\sum\limits_{m \in \mathcal M} {{c_{m,i,j}}}  \le H,\forall i \in \mathcal N,\forall j \in \mathcal N},\\
&&& {C2:\sum\limits_{i \in \mathcal N} \sum\limits_{j \in \mathcal N} {{c_{m,i,j}}} \le V, \forall m \in \mathcal M},\\
&&& {C3:{c_{m,i,j}} \in \{ 0,1\} ,\forall m \in \mathcal M,\forall i \in \mathcal N,\forall j \in \mathcal N},\\
&&& {C4:{R_{\sec ,m,i,j}}({c_{m,i,j}},{p_{m,j}}) \ge {R_{\min }}},\\
&&&\forall m\in \mathcal M,\forall i \in \mathcal N,\forall j \in \mathcal N, \\
&&& {C5:\sum\limits_{m \in \mathcal M} {{\sum\limits_{j \in \mathcal N} {{p_{m,j}}} } }  \le P_s},\forall m\in \mathcal M,\forall j \in \mathcal N,\\
&&& {C6:{p_{m,j}} \ge 0,\forall m \in \mathcal M,\forall j \in \mathcal N}.
\end{aligned}\label{eq:constraints weighted Sum rate maximization problem formulation}
\end{equation}

Constraints (C1)-(C2) restrict that each SC pair can occupy at most $H$ user pairs and each user pair can be assigned to at most $V$ SC pairs, separately; Constraints C3 ensures user pair scheduling variables to be binary. Constraints C4 ensures the QoS for each user pair, which requests secure data rate for each user pair must be larger than the minimum user pair data rate $R_{\min }$; Constraints (C5)-(C6) constraint that power variables satisfy transmitting power of the RS; The optimization problem is a non-convex optimization problem and an NP-hard problem.

The achievable secrecy energy efficiency affected by power allocation in the BC phase. The power allocation for user $A_m$ and on $SC_j$ denoted as

\begin{equation}
\begin{aligned}
{p_{{A_m},j}} = {p_n}\frac{{{{({G_{{A_m},j}})}^{ - \lambda }}}}{{\sum\limits_{m = 1}^M {{{({G_m}_{,j})}^{ - \lambda }}} }}
\end{aligned}\label{eq: FTPA power allocation}
\end{equation}
where $\lambda $ is a decay factor. When $\lambda  = 0$, it corresponds to equal power allocation among the allocated users. When $\lambda$ increases, it reflects more power is allocated to the user pair with poorer CRNN.

If we use constraint $\sum {_{m\in \mathcal M}{p_{m,j}}}  \le {P_s}/N,\forall m\in \mathcal M,\forall j \in \mathcal N$ to replace constraints C2 and C5, then the optimization problem is transformed into a closed-form optimal problem, which is easy to handle. Energy efficiency problem can be rewritten as
\begin{equation}
\begin{aligned}
\mathop {\max }\limits_{{c_{m,i,j}},{p_{m,j}}} \sum\limits_{m\in \mathcal M} {\sum\limits_{i \in \mathcal N} {\sum\limits_{j \in \mathcal N} {{\eta _{E}}({c_{m,i,j}},{p_{m,j}})} } }
\end{aligned}\label{eq:weighted Energy efficiency maximization problem formulation for power allocation}
\end{equation}

\begin{equation}
\begin{aligned}
& \text{subject to}
&&C1:\sum\limits_{m \in \mathcal M} {{\sum\limits_{j \in \mathcal N} {{p_{m,j}}} } }  = {P_s},\\
&&& {C2:{p_{m,j}} \ge 0,\forall m \in \mathcal M,\forall j \in \mathcal N}\\
&&& {C3:{R_{\sec ,m,i,j}}({c_{m,i,j}},{p_{m,j}}) \ge {R_{\min }}},\\
&&&\forall m\in \mathcal M,\forall i \in \mathcal N,\forall j \in \mathcal N, \\.
\end{aligned}\label{eq:constraints weighted Energy efficiency maximization problem formulation for power allocation}
\end{equation}

\section{Energy Efficient Subcarrier Matching Scheme For NOMA}\label{sec:Energy Efficient Subcarrier Matching Scheme For NOMA}

\subsection{Subcarrier Matching Problem Formulation}\label{subsec:Subcarrier Matching Problem Formulation}
We first introduce the concepts of matching game, preferred matched pair, preferred matching. Considering the set of user pairs and the set of SCs as two disjoint sets of players aiming to maximize their own energy efficiency, formally presented as.

Definition 1: (Two-sided Matching) \cite {Two-Sided Matching1992} Consider two disjoint sets, the user pairs \(\mathcal{M}=\{1,\ldots,M\}\), the SCs \(\mathcal{N}=\{1,\ldots,N\}\), a many-to-many mapping $\Phi$, such that for every $m \in \mathcal M$ and $S{C_i} \in \mathcal N$.
\begin{equation}
\begin{array}{*{20}{l}}
{1)\Phi (m) \subseteq \mathcal N,\Phi (S{C_i}) \subseteq \mathcal M};\\
{2)|\Phi (S{C_i})| \le H,|\Phi (m)| \le V };\\
{3)S{C_i} \in \Phi (m),m \in \Phi (S{C_i})}.\\
\end{array}
\end{equation}

Condition 1) implies that each user pair is matched with a set of SC pairs and each SC pair is matched with a set of user pairs. Condition 2) states that each SC pair can occupy at most ${H}$ user pairs, and each user pair can be assigned to at most ${V}$ SC pairs. To better describe the operation process of each player, Condition 3) means user $m$ and $SC_i$ are matched with each other.

Definition 2: (Preferred Match Pair) Given any two subcarriers $S{C_i},S{C_{i'}} \in \mathcal N,i \ne i'$, any one user pair $m$ and two matchings $\Phi $, $\Phi '$, $S{C_i} \in \Phi (m),S{C_{i'}} \in \Phi (m)$, if ${E_{m,i}}(\Phi ) > {E_{m,i'}}(\Phi ')$ implies that user pair ${m}$ prefers $S{C_i}$ in $\Phi $ to $S{C_{i'}}$ in $\Phi '$. Similarly, given any two user pairs ${m},{m'} \in \mathcal M,{m} \ne {m'}$, and two matchings $\Phi $, $\Phi '$, $m = \Phi (S{C_i}),m' = \Phi '(S{C_i})$, if ${E_{m,i}}(\Phi ) > {E_{m',i}}(\Phi ')$ implies that $S{C_i}$ prefers the user pairs ${m}$ to ${m'}$.

Since many-to-many matching is hard to achieve stable matching, we introduce the notion of switch matching as below.

Definition 3: (Preferred Matching) Given a matching $\Phi $ with $S{C_i} \in \Phi ({m}),S{C_j} \in \Phi ({n})$, and  $S{C_i} \notin \Phi ({n}),S{C_j} \notin \Phi ({m})$, if ${E_i}(m) > {E_j}(n)$, there exists $\Phi _{n,j}^{m,i}$;

\begin{equation}
\begin{array}{*{20}{l}}
{S{C_i} \in \Phi _{n,j}^{m,i}(m),S{C_j} \in \Phi _{n,j}^{m,i}(n),}\\
{S{C_i} \notin \Phi _{n,j}^{m,i}(n),S{C_j} \notin \Phi _{n,j}^{m,i}(m)}
\end{array}
\end{equation}
$\Phi _{n,j}^{m,i}$ is called a preferred matching. Two user pairs in the same subset exchange their matches in the opposite subset while other matches remain unchanged. Note that if a preferred matching is approved, then at least one player's data rates will increase, and the achievable rates of any player involved will not decrease at the same time.

\subsection{Subcarrier Assignment Algorithm For NOMA}\label{subsec:Subcarrier Assignment Algorithm For NOMA}
We formulate two SC assignment algorithms (SCAS-1 and SCAS-2). In SCAS-1, we assume that a larger CRNN of the SC has a higher priority to select user pairs. The preferred matching phase in SCAS-2, the RS keeps searching for two user pairs to form a match pair, then executes the preferred matching and updates the current matching if satisfied conditions. The iterations stop until no user pairs can form a new match pair. The algorithms are described in detail in Table I and Table II as follow.

\begin{algorithm}[!ht]
\caption{Subcarrier Assignment Scheme (SCAS-1)}
\begin{algorithmic}[1]
\STATE  Based on the CSI of each SC, the RS allocates the transmission of power equally to each SC;
\STATE  Initialize all of unmatched user pairs  and unmatched SCs;
\nonumber\REPEAT
\IF     {$ |\Phi (S{C_i})| \le {H} $}
\STATE  $S{C_i}$ selects its most preferred unmatched user pair by using CRNNs;
\ENDIF
\IF     {$ |\Phi (S{C_i})| = {H} $}
\STATE  Set the proportional factor of power for each user pair according to \eqref{eq: FTPA power allocation};
\STATE  For any two SCs $S{C_i}$ and $S{C_j}$ select any two user pairs $m$ and $n$, respectively. $S{C_i} \in \Phi (m),S{C_j} \in \Phi (n), S{C_i} \notin \Phi (n), S{C_j} \notin \Phi (m)$;
\WHILE  {${E_i}(m) > {E_j}(n)$}
\STATE  Execute preferred matching $S{C_i} \in \Phi (n), S{C_j} \in \Phi (m)$, $S{C_i} \notin \Phi (m), S{C_j} \notin \Phi (n)$;
\STATE  Update all user pairs' energy efficiency;
\ENDWHILE
\ENDIF
\UNTIL Convergence
\end{algorithmic}\label{Algorithm SCAS-1}
\end{algorithm}

\begin{algorithm}[!ht]
\caption{Subcarrier Assignment Scheme (SCAS-2)}
\begin{algorithmic}[1]
\STATE  Based on the CSI of each SC, the RS allocates the transmission of power equally to each SC;
\STATE  Initialize all of unmatched user pairs and unmatched SCs;
\nonumber\REPEAT
\IF     {$|\Phi (S{C_i})| \le {H}$ and $|\Phi ({m})| \le {V}$}
\STATE  All of the SCs and user pairs are matched with each other arbitrarily;
\ENDIF
\IF     {$ |\Phi (S{C_i})| = {H} $}
\STATE Set ${E_{\sec ,\max }} = {E_{\sec ,total}}(\Phi )$;
\WHILE{$\ell < {L_m}$}
\STATE Two user pairs $(m,n)$ and SCs $(S{C_i},S{C_j})$ are selected with $S{C_i} \in \Phi (m),S{C_j} \in \Phi (n)$ ,$S{C_i} \notin \Phi (n),S{C_j} \notin \Phi (m)$;
\WHILE{${E_{\sec ,tatol}}(\Phi _{n,j}^{m,i}) > {E_{\sec ,\max }}$}
\STATE Execute preferred matching $\Phi _{n,j}^{m,i}$;
\STATE Set ${E_{\sec ,\max }} = {E_{\sec ,tatol}}(\Phi _{n,j}^{m,i})$;
\STATE $\ell  = \ell  + {\rm{ 1}}$;
\ENDWHILE
\ENDWHILE
\ENDIF
\UNTIL Convergence
\end{algorithmic}\label{Algorithm SCAS-2}
\end{algorithm}

\section{Energy Efficient Power Allocation Scheme For NOMA}\label{sec:Energy Efficient Power Allocation Scheme For NOMA}

As mentioned in Section \ref{sec:Energy Efficient Subcarrier Matching Scheme For NOMA}, we investigate the SC assignment scheme in the two-way relay NOMA network. In order to further improve the system energy efficiency, we design user pairs' power allocation algorithm instead of equal power allocation. In this section, we introduce GP programming approach and discuss its effects of different power proportional factor on the energy efficiency of the system.

\subsection{Energy Efficient Power Allocation Algorithm For NOMA}\label{subsec:Algorithm Design}

The objective function in \eqref{eq:weighted Energy efficiency maximization problem formulation for power allocation} is non-convex. We introduce the parameter transformation to avoid high complexity of the solution. We assume that the proportion of the power is assigned on SC $i$ in the MA phase and on SC $j$ in the BC phase, denoted by $u_{i,j}$. We can formula the non-convex optimization problem as GP \cite{GPOptandEng2007}. Therefore, the energy efficiency maximization for power allocation problem can rewrite as
 \begin{equation}
\begin{array}{*{20}{l}}
{u_{E}} = \max \frac{{{R_{\sec ,m,i,j}}({c_{m,i,j}},{p_{m,j}})}}{{{P_s}({c_{m,i,j}},{p_{m,j}})}}
\end{array}\label{eq: }
\end{equation}

 \begin{equation}
\begin{array}{*{20}{l}}
\min {u_{i,j}}{P_s}({c_{m,i,j}},{p_{m,j}}) - {R_{\sec ,m,i,j}}({c_{m,i,j}},{p_{m,j}})
\end{array}\label{eq:final Energy efficiency}
\end{equation}

\begin{equation}
\begin{aligned}
& \text{subject to}
&& C1:\sum\limits_{m \in \mathcal M} {{\sum\limits_{j \in \mathcal N} {{p_{m,j}}} } }  = {P_s},\\
&&& {C2:{p_{m,j}} \ge 0,\forall m \in \mathcal M,j \in \mathcal N}\\
&&& C3:{R_{\sec ,m,i,j}}({c_{m,i,j}},{p_{m,j}}) \ge {R_{\min }},\\
&&&\forall m\in \mathcal M,\forall i \in \mathcal N,\forall j \in \mathcal N \\
&&& C4:\sum\limits_{i \in \mathcal N} {\sum\limits_{j \in \mathcal N} {{u_{i,j}}} }  = 1.
\end{aligned}\label{eq:constraints weighted Energy efficiency maximization problem formulation for power allocation}
\end{equation}

In the following, an iterative resource allocation algorithm for power allocation can be proposed. Due to the objective function is linear and the constraints are convex, we can utilize interior point methods to solve the global-optimal problem\cite{Cambridge2004}\cite{NOMAIEEETrans2016}. The secrecy energy efficiency significantly improves for each iteration in algorithm 3.

\begin{algorithm}[!ht]
\caption{ A Novel Power Allocation Algorithm }
\begin{algorithmic}[1]
\STATE  Based on the CSI of each SC, the RS allocates the transmission of power equally to each SC;
\STATE Initialize the maximum tolerance $ \varepsilon $ and the number of iterations $\ell$ and the maximum number of iterations $L_m$;
\WHILE  {$|{R_{\sec ,m,i,j}} - {u_{E}}{P_s}| > \varepsilon$ or $\ell \le {L_m}$}
\STATE Update $p$ by solving the formulated problem in \eqref{eq:final Energy efficiency} and \eqref{eq:constraints weighted Energy efficiency maximization problem formulation for power allocation} using the interior point methods;
\STATE $\ell = \ell + 1$;
\ENDWHILE
\end{algorithmic}\label{Power Allocation Algorithm 3}
\end{algorithm}

\subsection{Stability, Convergence and Complexity}\label{subsec:Stability, Convergence and Complexity}
We give remarks on the stability, convergence and complexity for proposed SSPA schemes.

1)Stability

When no user pair ${m} \in \mathcal M$ can find another user pair ${n} \in \mathcal M$ to match with each other, we regard the proposed SSPA convergence as the best choice for current matching. There is any user pair cannot improve its utility by using to change its matches. Hence, the terminal matching ${\Phi ^*}$ is two-sided exchange stable matching, which guarantees the proposed SSPA schemes to remain stable.

2) Convergence

Convergence of SSPA algorithms depend on preferred matching and power allocation by using interior point methods.

1) After a number of match operations for one time power allocation:

We assume that the preferred matching are $(Am,Bm)$ and $(An,Bn)$ with ${\Phi _\ell } = {\Phi _{\ell  - 1}}_{nj}^{mi}$. We ensure that the utility of $S{C_p}$ and $S{C_q}$ satisfies ${E_{S{C_p}}}({\Phi _\ell }) \ge {E_{S{C_p}}}({\Phi _{\ell  - 1}})$ and ${E_{S{C_q}}}({\Phi _\ell }) \ge {E_{S{C_q}}}({\Phi _{\ell  - 1}})$ after each preferred matching. Hence, the total secrecy energy efficiency increase after each match operation $\ell $. There exists a preferred matching after which the total secrecy energy efficiency stops increasing for one time power allocation.

2) After a number of power allocation for the SSPA algorithms:

The total energy efficiency will increase after a number of iterations. Due to the limited spectrum resource the total secrecy energy efficiency has an upper limit. Hence, the total secrecy energy efficiency will stop increasing after a limited number of iterations. Therefore, the proposed SSPA algorithms for secure resource allocation is guaranteed to converge.

3) Complexity

The complexity of the proposed SSPA-1 algorithm is $O(PSM{H}{V}(N - {V}){M^2})$. When $ |\Phi (S{C_i})| \le {H} $. The complexity mainly produces in the process of using the CRNN, which is O($M^2$). For user pair $m$, there exist $V(N - V)$ possible combinations of $S{C_i}$ and $S{C_j}$ in $\Phi _{n,j}^{m,i}$ need to be considered. For the $S{C_i}$, the preferred matching $\Phi _{n,j}^{m,i}$ with $m$ has ${H}{V}\left( {N{\rm{ }} - {\rm{ }}{V}} \right)$ possible combinations. There are $\frac{1}{2}M{H}{V}\left( {N - {V}} \right)$ preferred matchings need to be considered for $M$ user pairs in each iteration. The number of total match iterations is $S$ and power allocation iterations is $P$. The computational complexity of SSPA-1 can be presented by $O(PSM{H}{V}(N - {V}){M^2})$. Similar analysis can be performed for the proposed SSPA-2 algorithm and the complexity of which is $O(PSM{H}{V}(N - {V}){N^{H + 1}})$.

\section{Simulation Results and Discussion}
In this section, we evaluate the performance of the proposed SSPA schemes with both SCAS-1 and SCAS-2 applied, and compare its performance with a random allocation scheme (RA-NOMA). We assume that two adjacent users are considered as a user pair, which selects the same SC. Each SC can be assigned to at most ${H} = 3$  user pairs, and each user pair can occupy at most ${V} = 4$  SCs.  In the RA-NOMA scheme, the SCs is randomly allocated to the user pairs satisfying ${H}\le  3$ and ${V} \le  4$. For the simulations, the total of RS’s peak power Ps is 46dBm, system bandwidth is 4.5MHz and the transmit power for each user is ${P_{A_m}} = 300mW, {P_{B_m}} = 300mW$ on the uplink. We assume that noise power spectral density is -150 dBm/Hz, circuit power consumption ${P_c} = 1$dB and eavesdropper is allocated at a distance of 500 m from the RS, if there is no special instructions. Pass loss functions can be obtained by hata urban propagation model \cite{3GPPTR2014}. The coverage radius of the RS is $r = 30$ m and user pairs are evenly distributed in a circle around the central RS. Considering the computational complexity, we assume that there are 10 SCs in the NOMA wireless network.

Fig. 2 shows the cumulative distribution function (C.D.F.) of the total number of iterations for the SCAS-1 to convergence. To evaluate the performance of the proposed SCAS-1 scheme, we adopt \eqref{eq: FTPA power allocation} power allocation scheme for each user. Note that the random variable $\mathcal{Y}$  is the total number of match numbers required for the SCAS-1. With the number of the user pairs increasing, the speed of convergence becomes slower. In addition, the proposed SCAS-1 converges within 2-10 match operations and further reflecting the relatively low computational complexity.
\begin{figure}[t]
        \centering
        \includegraphics*[width=80mm]{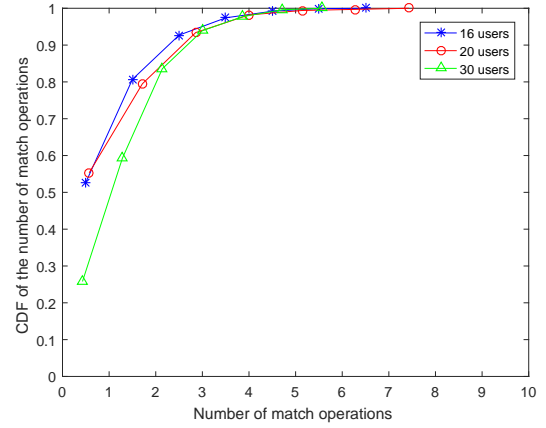}
        \caption{Number of match operations vs. the number of users.}
        \label{fig:2}
\end{figure}

Fig. 3 shows secrecy energy efficiency performance vs. the number of match operations in the SCAS-2 scheme. When users become larger, the number of match operations become higher due to more user pairs have the opportunity to be serviced by the RS. From Fig. 3, we can see energy efficiency increases with the match operation number increasing within 40 match operations. The energy efficiency closes to a relatively stable level when match operation number over 40 match operations which implies the proposed SCAS-2 scheme also has a low complexity.
\begin{figure}[t]
        \centering
        \includegraphics*[width=80mm]{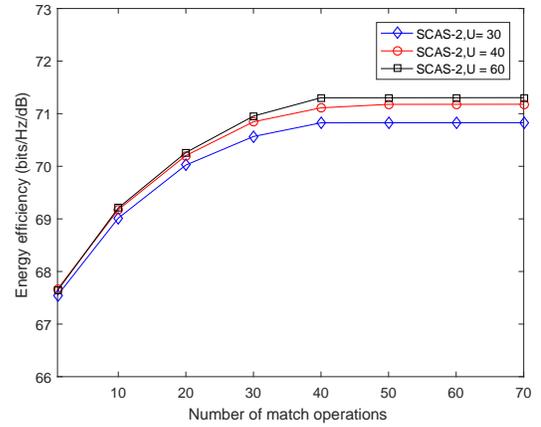}
        \caption{C.D.F. of the number of match operations in SCAS-2.}
        \label{fig:3}
\end{figure}

Fig. 4 illustrates the secrecy energy efficiency performance vs. the number of iterations for the proposed algorithm 3. We can see that the proposed SSPA algorithms based power allocation algorithm (Algorithm 3) takes at most 10 iterations to converge. The energy efficiency goes up sharply with 1 iteration, and then it closes to a relatively stable level when the number of iterations over 1 time. However, although the SSPA-2 scheme show a higher iteration than SSPA-1 scheme and RA-NOMA scheme, it provides a higher secrecy energy efficiency than the other schemes.
\begin{figure}[t]
        \centering
        \includegraphics*[width=80mm]{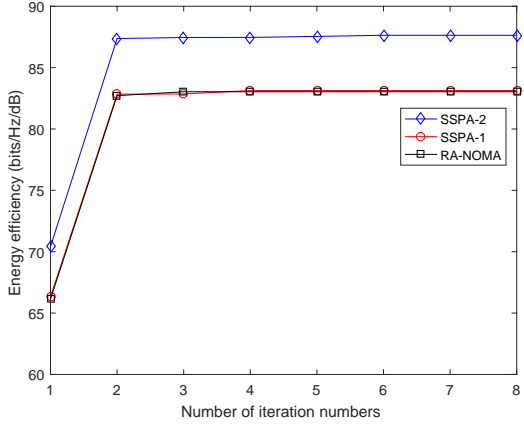}
        \caption{Secrecy energy efficiency vs. number of iterations.}
        \label{fig:4}
\end{figure}

Fig. 5 illustrates the secrecy energy efficiency performance vs. ${P_{{A_m}}}/{\sigma ^2}$ for the two SSPA proposed schemes and RA-NOMA scheme. As the ${P_{{A_m}}}/{\sigma ^2}$ grows, the secrecy energy efficiency continues to increase, but the growth rate slows down. When ${P_{{A_m}}}/{\sigma ^2}$ over 155 db, the cochannel interference seriously affected the performance of the system the RA-NOMA scheme is worse than the OFDMA scheme. From Fig. 5, we can also see that both SSPA-1 scheme and SSPA-2 scheme have better performance than RA-NOMA scheme, proving that SSPA schemes effectively improve the system's secrecy energy efficiency. Meanwhile, since SSPA-2 provides more freedom in the SC allocation than the randomly predefined user pairs in the SSPA-1, the SSPA-2 scheme thoroughly outperforms the SSPA-1 scheme.
\begin{figure}[t]
        \centering
        \includegraphics*[width=80mm]{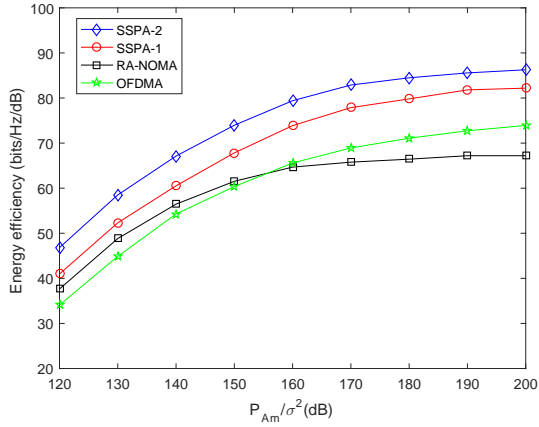}
        \caption{Secrecy energy efficiency vs. ${P_{{A_m}}}/{\sigma ^2}$.}
        \label{fig:5}
\end{figure}

Fig. 6 depicts the secrecy energy efficiency of the proposed SSPA schemes and RA-NOMA scheme vs. circuit power consumption of each SC. The circuit power consumption for each SC is set from 0.1 dB to 1.5 dB. As shown in this figure, we can be observed that the secrecy energy efficiency decreases with a grow of circuit power consumption. The main factor is that more energy is consumed by the circuit, the energy used for signal transmission will be greatly reduced. Thus, as expected, the circuit power consumption can degrade the secrecy energy efficiency performance.

\begin{figure}[t]
        \centering
        \includegraphics*[width=80mm]{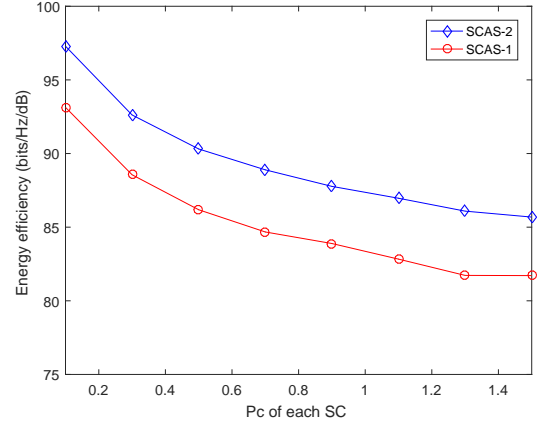}
        \caption{Secrecy energy efficiency vs. Pc of each SC.}
        \label{fig:6}
\end{figure}

Fig. 7 depicts the secrecy energy efficiency performance vs. ${\sigma ^2}$ for the two SSPA proposed schemes and RA-NOMA scheme with CJ. Under the CJ scenario, we propose generating AN at the RS for further enhancing the secrecy performance. The performance of the proposed SSPA-2 scheme for the NOMA wireless network achieves higher energy efficiency than that of SSPA-1 scheme as well as RA-NOMA scheme. Fig. 7 demonstrates that the secrecy energy efficiency is remarkably deteriorated by the additive white Gaussian noise. Therefore, ${\sigma ^2}$ increases, as a result, the secrecy energy efficiency decrease. The secrecy energy efficiency obtained from the proposed SSPA-1 scheme drops down fast for ${\sigma ^2}$ over -90 dBm, and less so fast ${\sigma ^2}$ within -90 dBm.

\begin{figure}[t]
        \centering
        \includegraphics*[width=80mm]{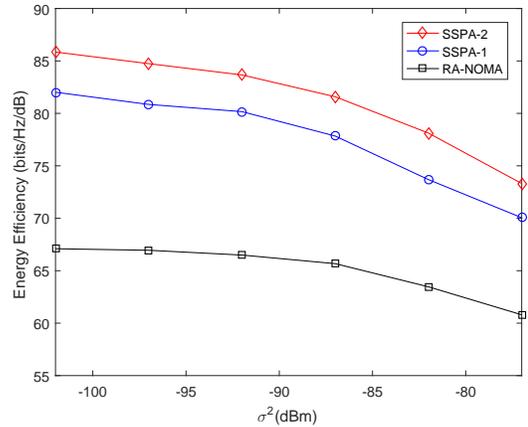}
        \caption{Secrecy energy efficiency vs. ${\sigma ^2}$.}
        \label{fig:7}
\end{figure}

Fig. 8 shows the secrecy energy efficiency performance vs. ${\sigma ^2}$ for $N = 6$, $N = 8$, $N = 10$ in SSPA-1 with CJ. As observed in Fig. 8, The energy efficiency decreases sharply with the ${\sigma ^2}$ increases. Obviously, when secrecy energy efficiency increases with the number of SCs $N$ grows, due to the more user pairs be allocated over SCs.
\begin{figure}[t]
        \centering
        \includegraphics*[width=80mm]{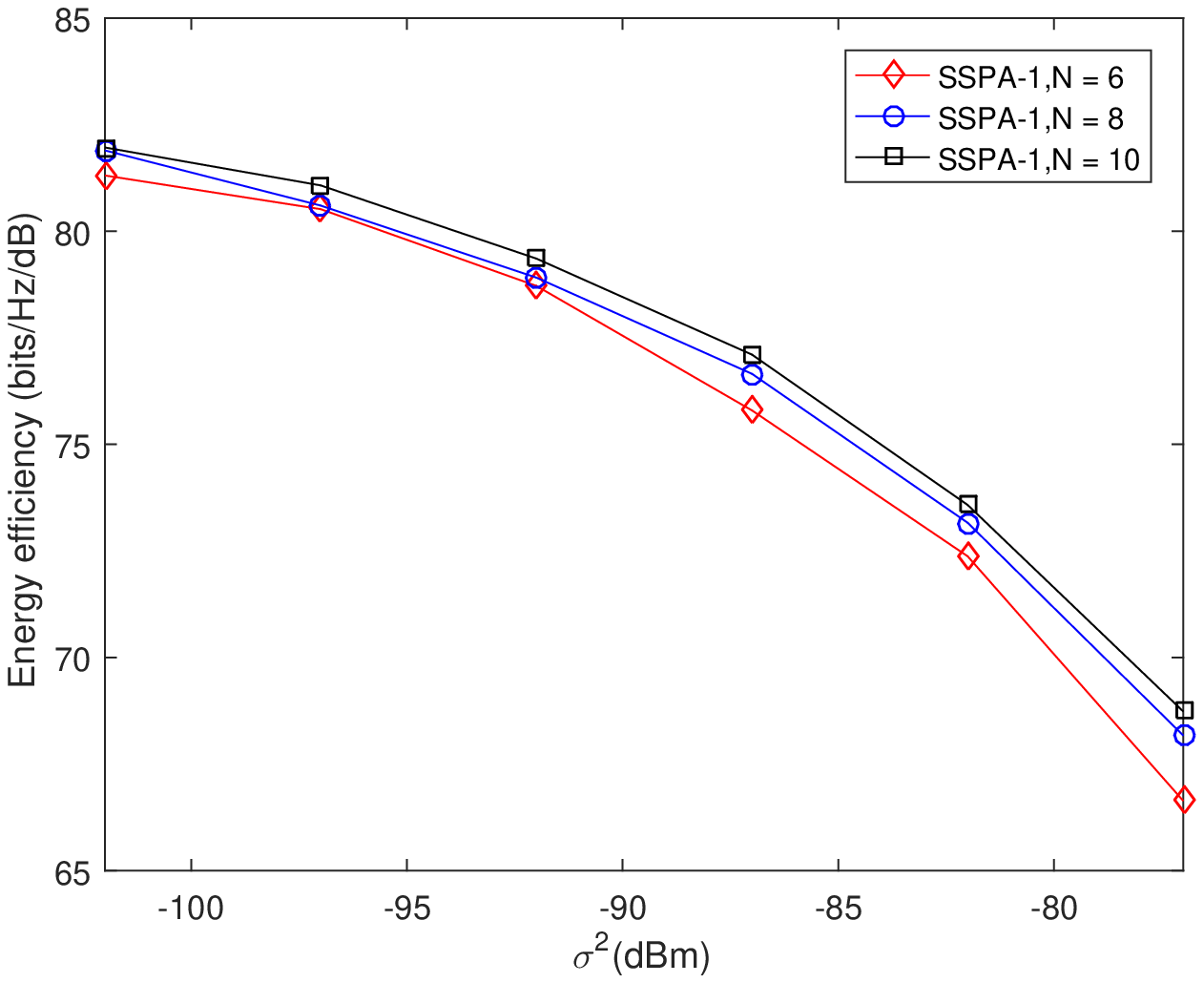}
        \caption{Secrecy energy efficiency vs. ${\sigma ^2}$.}
        \label{fig:8}
\end{figure}

Fig. 9 shows the secrecy energy efficiency performance vs. ${\sigma ^2}$ for different value of user pairs in SSPA-2 with CJ. Similar to Fig. 7 and Fig. 8 show that the secrecy energy efficiency in NOMA wireless network deceases when ${\sigma ^2}$ increased. It can be observed from the figure, the more number of user pairs in the NOMA wireless network is, the better of the performance is obtained. The main reason is that, as the number of the total SC is fixed as N, with the increase of the number of user pairs, the more diversity gains over SCs.

\begin{figure}[t]
        \centering
        \includegraphics*[width=80mm]{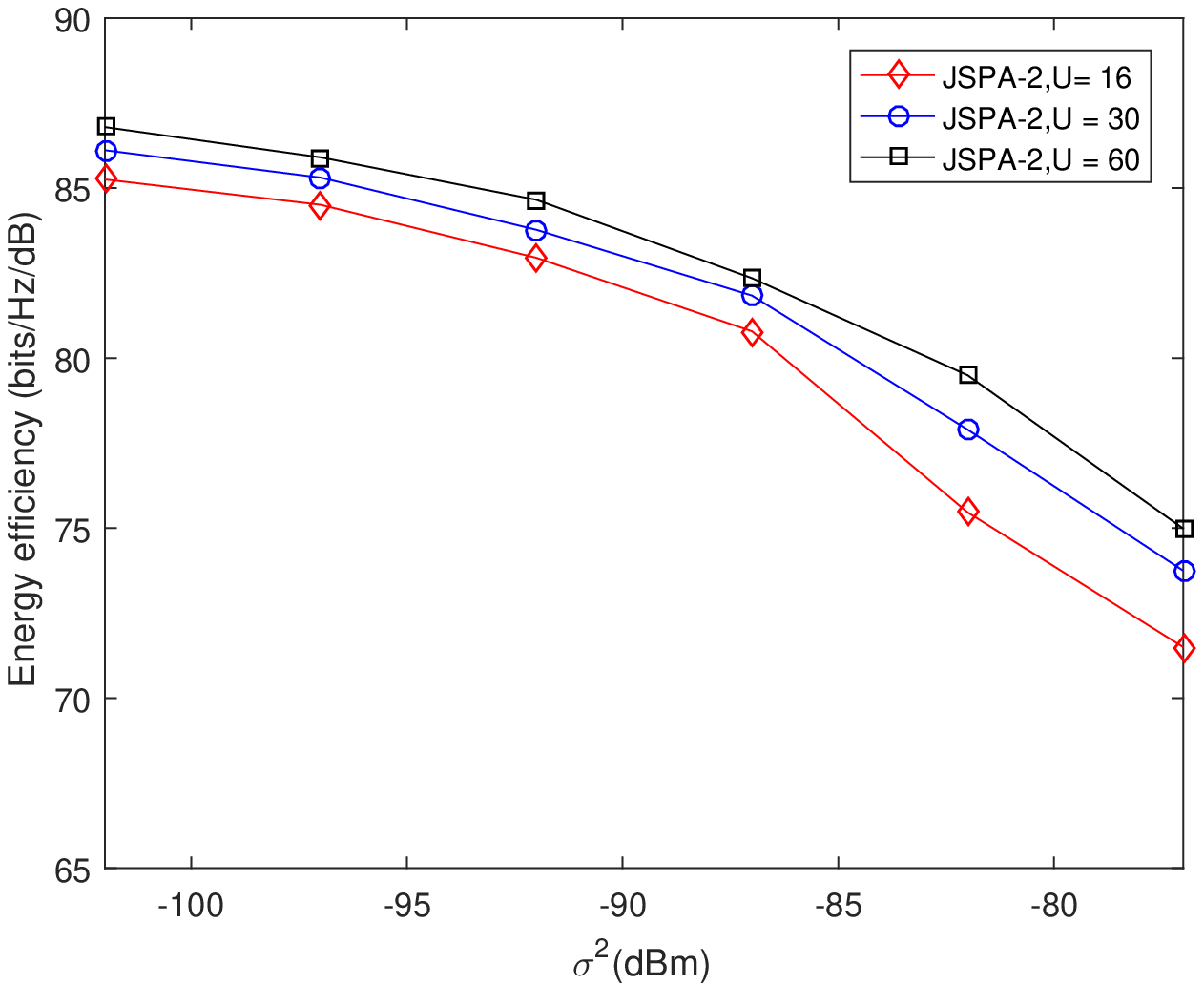}
        \caption{Secrecy energy efficiency vs. ${\sigma ^2}$.}
        \label{fig:9}
\end{figure}

\section{Conclusion}
In this paper, we investigated the secure SC assignment and power allocation for the NOMA two-way relay wireless networks in the presence of an eavesdropper without and with CJ. The proposed SSPA algorithms with SCAS applied properly allocate resources to user pairs, and the performance of secrecy energy efficiency of the system can be significantly improved than the RA-NOMA scheme. Moreover, the SSPA-2 scheme thoroughly outperforms the SSPA-1 scheme.

%

\end{document}